\documentstyle[prl,aps,epsf]{revtex}
\newcommand{\be}{\begin{equation}}
\newcommand{\ee}{\end{equation}}
\newcommand{\bea}{\begin{eqnarray}}
\newcommand{\eea}{\end{eqnarray}}
\begin{document}
\title{
Gravitating (bi-)sphalerons}
\author{{\large Y. Brihaye and M. Desoil} }
\address{\small 
Facult\'e des Sciences, Universit\'e de Mons-Hainaut,
B-7000 Mons, Belgium }
\address{ }

\date{\today}

\maketitle
\begin{abstract}
The standard model of electroweak interactions
is minimally coupled to gravity and the response
of the spherically symmetric solutions -the sphaleron
and the bisphaleron- to gravity is emphasized.
For  a given value of the Higgs mass $M_H$, several 
branches of solutions exist which terminate into 
cusp-catastrophy at some ($M_H$-depending) critical value
of the parameter $\alpha$ defined by the ratio of the 
vector-boson mass to the Planck mass.
A given branch either bifurcates from another one at an
intermediate value of $\alpha$ or persists in the limit
$\alpha \rightarrow 0$ where it terminates into a flat
sphaleron or bisphaleron or into a Bartnik-McKinnon
solution. These bifurcation patterns are studied in some details.
\end{abstract}

\section{Introduction}

The coupling of non-abelian field theories to Einstein gravity
constitutes natural extensions of the Einstein-Maxwell equations.
One of the surprising issues of such possibilities has been the 
discovery of regular, finite energy classical solutions in the
Einstein-Yang-Mills equations~: the series of Bartnik-Mckinnon
\cite{bm}   solutions.

On the other hand the Weinberg-Salam model theory, which couples
the SU(2)$\otimes$U(1) Yang-Mills theory  to a doublet of complex 
Higgs fields emerges more and more as the theory of the unified weak 
and electromagnetic forces. 
This particular lagrangian is a member of the family of Yang-Mills-Higgs 
models among which the Georgi-Glashow model is another distinguished
example.

One interesting property of the Weinberg-Salam lagrangian is that it 
admits an unstable,
finite energy classical solution called the sphaleron
\cite{km}.
Generally, sphaleron solutions are expected to play
a role in the understanding of baryon non conserving phenomenon
which are allowed to take place in the Weinberg-Salam model
\cite{thooft}.
When the parameter determining the mass of the Higgs field
increases, additional solutions, the bisphalerons, bifurcate
from the sphaleron \cite{bk,yaffe}. 
This feature seems to be related to the underlying
non-linear character of the classical equations and to
the spontaneous breakdown of the symmetry \cite{ms,bgkk,bt}.

It is therefore natural to couple the Weinberg-Salam 
model to the Einstein-Hilbert gravitational lagrangian and to study
the response to gravity of the classical solutions 
available in the flat limit.
This problem has been partly investigated 
in \cite{greene} and more recently in  a  report on
the aspects of Einstein-Yang-Mills-Higgs equations \cite{volkov}
but it was not treated in details. 
In particular the response of the bisphaleron solution to
gravity was not investigated.
In this paper, we reconsider the classical equations of the 
Einstein-Weinberg-Salam model for  spherically symmetric 
fields (we assume the Weinberg angle, related to the U(1) part
of the gauge group, to be zero).
Then the gravitating sphaleron and bisphaleron are constructed
and assemble into branches of solutions which evolve with $\alpha$,
the ratio of the vector-boson mass to the Planck mass.
The evolution of the different solutions as functions of $\alpha$
and $M_H$ are studied in details.

\section{The equations}

We consider the gauge theory for an SU(2)-Higgs doublet
minimally coupled to the Einstein-Hilbert gravitational
lagrangian~:
\be
\label{lagrangien}
L   = \sqrt{-g} [ L_G + L_M]
\ee
with 
\be
       L_G = \frac{1}{16 \pi G} R
\ee
\be
L_M =  -\frac{1}{4} F_{\mu \nu}^a F^{a \mu \nu}
       + (D_{\mu} \Phi)^{\dagger} (D^{\mu} \Phi)
       - \frac{\lambda}{4} (\Phi^{\dagger} \Phi - \frac{v^2}{2})^2
\ee
and with the usual definitions for the fields 
strenghts and covariant derivatives
\be
    F_{\mu \nu}^a = \partial_{\mu} A_{\nu}^a 
                    - \partial_{\nu} A_{\mu}^a
        + g \epsilon_{abc} A_{\mu}^b A_{\nu}^c
\ee                 
\be
    D_{\mu} \Phi = \partial_{\mu} \Phi + g (A_{\mu}^a \sigma_a) \Phi
\ee
The matter part $L_M$ of this Lagrangian approximates the
Weinberg-Salam model of electroweak interactions in the limit
of vanishing Weinberg angle $\theta_W$ (i.e. the
gauge group SU(2)$\times$U(1) is restricted to SU(2)).

Here we will study the classical, spherically symmetric 
solutions of  the Lagrangian $L$. In this purpose, we employ the
Schwarzschild-like coordinates for the metric
\begin{equation}
\label{metric}
ds^2=
  -A^2N dt^2 + N^{-1} dr^2 + r^2 (d\theta^2 + \sin^2\theta d\phi^2)
\ , \end{equation}
and we introduce, as usual, the mass function $m(r)$ by means of
\begin{equation}
N(r)=1-\frac{2m(r)}{r} 
\ . \label{n} \end{equation}
Then we use
the standard spherically symmetric ansatz for the spatial 
components of the gauge field (fields are static and
$A_0=0$)  and for the
Higgs fields (the notations of \cite{akiba} are used)
\begin{equation}
A_i^a = \frac{1-f_A(r)}{gr} \epsilon _{aij} \hat x_j
       + \frac{f_B(r)}{gr} (\delta_{ia} - \hat x_i \hat x_a)
       + \frac{f_C(r)}{gr} \hat x_i \hat x_a
\ , \end{equation}
and 
\begin{equation}
\Phi_1 = 0 \ \ , \ \ \Phi_2 = 
\frac{v}{\sqrt 2} \bigl[ H(r) + i K(r) (\hat x^a \sigma_a) \bigr]
\ . \end{equation}

It is well known that this ansatz for the matter fields
is plagued with a residual gauge symmetry. Along with \cite{akiba},
we  fix this freedom by imposing the axial gauge
\be
      x^i A_i = 0 \ \ \  \Rightarrow \ \ \ f_C(r) = 0 \ \ .
\ee
It is also convenient to introduce the dimensionless coordinate $x$
and mass function $\mu$ defined by
\begin{equation}
x = g \frac{v}{\sqrt 2} r \ , \ \ \ \mu=g \frac{v}{\sqrt 2} m
\ , \label{xm} \end{equation}
as well as the dimensionless coupling constants $\alpha$, $\epsilon$
\begin{equation}
\label{alpha}
\alpha^2 = 4 \pi G \frac{v^2}{2} \ \ \ , \ \ \ 
\epsilon = \frac{\lambda}{g^2} = \frac{1}{2} (\frac{M_H}{M_W})^2 \ ,  
\end{equation}
where $G$ is Newton's constant, 
$v$ is the Higgs field expectation value,
 $M_H$ is the Higgs boson mass and $M_W$ is the gauge boson mass.
If finite, the quantity $\mu(\infty)$ defines the mass of the 
solution.

With these ansatz and definitions, it can be checked
after an algebra that the classical equations
of the Lagrangian (\ref{lagrangien}) are equivalent 
to the equations derived
from the following two-dimensional action
\be
    S = \int dt \ dx (- {\cal L} )
\ee
\be
{\cal L} = A [ \frac{1}{2}(N + x N' - 1) + \alpha^2 M ]
\ee
where the prime means the derivative with respect to $x$
and the quantity $M$ is defined by
\bea
  &M   &= N V_1 + V_2 \\
  &V_1 &= (f_A')^2 + (f_B')^2 + 2 x^2 ( (H')^2 + (K')^2)  \\
  &V_2 &=  \frac{1}{2 x^2} (f_A^2 + f_B^2 - 1)^2 
       + \epsilon  x^2 (H^2 + K^2 - 1)^2 \\ 
  &    &+ (H(f_A-1) + K f_B)^2 + (K(f_A+1) - H f_B)^2  \ .   
\eea

The classical equations then reduce to the 
following system of six  non-linear differential equations~:
\begin{eqnarray}
    \mu' &=& \alpha^2 M \\
    A'   &=& 2 A \alpha^2 \frac{1}{x} V_1 \\
 (A N f_A')' &=& \frac{1}{2} A (\frac{\partial V_2}{\partial f_A}) \\
 (A N f_B')' &=& \frac{1}{2} A (\frac{\partial V_2}{\partial f_B}) \\
 (x^2 A N H')' &=& \frac{1}{4} A (\frac{\partial V_2}{\partial H}) \\
 (x^2 A N K')' &=& \frac{1}{4} A (\frac{\partial V_2}{\partial K}) 
\end{eqnarray}

It is important to remark that the equations are still invariant
under the continuous global transformation
\bea
    & f_A + i f_B  \longrightarrow &({\rm exp} (2 i \Omega)) (f_A + i f_B) \nonumber \\
    & H + i K  \longrightarrow &({\rm exp}(i \Omega)) (H + i K)
\label{symmetry}
\eea
where $\Omega$ is a real constant. 
Since the regularity of the functional $V_2$ at $x=0$
clearly requires $f_A^2(0) + f_B^2(0) = 1$, we can fix the above
symmetry by chosing $f_A(0) = 1$,$f_B(0) = 0$.
This partly fix the boundary conditions
which will be discussed more completely in the next section.
Then, only a change of sign of the Higgs field can still be
chosen arbitrarily. 

Let us close this section by discussing the limit $\alpha \rightarrow 0$.
From the definition (\ref{alpha}) it is clear that the limit
of vanishing $\alpha$ can be considered in different ways.
If we keep $v$ fixed and let $G \rightarrow 0$ we obtain the flat
limit (gravity decouples). 
The appropriate parameter
which defines the classical energy of the solution is 
\be
       E_c = {\rm lim}_{x \rightarrow \infty} \frac{1}{\alpha^2} 
\mu(x) \  .
\ee
It restitutes the physically-meaningful
classical energy of the flat (bi-)sphaleron
\cite{km,bk,yaffe}.
In order to study the equations in the limit 
$v \rightarrow 0$ and $G$ fixed, then it is necessary 
to rescale the radial variable $x$ and mass $\mu$ according to
\be
     y \equiv \frac{x}{\alpha} \ \ \ , \ \ \ 
     \rho \equiv \frac{\mu}{\alpha}
\ee
and to set $\alpha = 0$ afterwards in the equations. 
We then obtain
\bea
& \frac{d \rho}{dy} &= N ( (\frac{d f_A}{dy})^2 + (\frac{d f_B}{dy})^2 ) 
             + \frac{1}{2 y^2} (f_A^2 + f_B^2 - 1)^2 \\
& \frac{dA}{dy} &= 2 A \frac{1}{y} ((\frac{d f_A}{dy})^2 + (\frac{d f_B}{dy})^2 ) \\
& \frac{d}{dy}(A N \frac{d f_J}{dy}) &= A \frac{1}{y^2} f_J (f_A^2 + f_B^2 - 1) \ \ ,  \ \ J = A,B  .
\label{bmeq}
\eea
In particular, the degrees of freedom $H,K$ related
to the gauge field decouples and, setting $f_B=0$ in the
equations above,
the Bartnik-McKinnon (BM) equations \cite{bm}
are recovered.
In passing note that we have not succeeded in constructing 
solutions of (\ref{bmeq}) which are not related to the 
BM solution by mean of (\ref{symmetry}).

The corresponding energy is given by
\be
    E_{BM} = {\rm lim}_{y \rightarrow \infty} \rho(y) = 
             {\rm lim}_{x \rightarrow \infty} \frac {\mu(x)}{\alpha}
\ee

\section{Boundary conditions}

The regularity of the solution at the origin, the finiteness
of the mass $\mu(\infty)$ and the requirement that the metric (\ref{metric})
approaches the Minkowski metric for $x\rightarrow \infty$
lead to definite boundary conditions (BC) for the six radial
functions $\mu, A, f_A, f_B, H, K$.
As far as the metric functions are concerned we have to impose
\be 
\mu(0) = 0 \ \ , \ \ A(\infty) = 1
\ee
For the matter field equations two sets of BC
lead to regular and finite energy solutions.
\cite{bk}
\subsubsection{The sphaleron BC}
The flat sphaleron (and also the gravitational one) has
$f_B(x) = H(x) = 0$. The remaining functions have to obey
\bea
&f_A(0) = 1 \ \ \ \ , \ \ \ \ &K(0) = 0 \\
&f_A(\infty) = -1 \ \ \ \ , \ \ \ \ &K(\infty) = 1
\eea
\subsubsection{The bisphaleron BC}
The flat bisphalerons are characterized by the four 
non-trivial radial functions and we will see in the next
section that they are continuously deformed by gravity.
Taking into account the fixing of the global
symmetry (\ref{symmetry}), at the origin the functions have to obey
\bea
&f_A(0) = 1 \ \ \ \ , \ \ \ \ &f_B(0) = 0 \\
&H'(0) = 0 \ \ \ \ , \ \ \ \ &K(0) = 0 \\
\eea
In the limit $x\rightarrow \infty$, 
they have to approach constants in the following way
\bea
 &{\rm lim}_{x \rightarrow \infty} (f_A(x) + i f_B(x)) &= 
{\rm exp} (2 i \pi q) \\
 &{\rm lim}_{x \rightarrow \infty} (H(x) + i K(x)) &= 
{\rm exp} ( i \pi q)
\eea
So, the solutions of this type are  characterized by a real 
constant $q \in [0,1[$; this parameter
has to be determined numerically and
depends on $\epsilon$ and $\alpha$. 

\section{The solutions}

We first describe the solutions for $\epsilon = 0.5$,
a generic value of $\epsilon$ for which the sphaleron
is the unique solution of the flat equations.
The flat sphaleron is there for $\alpha = 0$, with an energy
$E_c \approx 3.64$, and gets continuously deformed for $\alpha > 0$.
In particular the function $N(x)$ develops a minimum which becomes
deeper while $\alpha$ increases. This branch of 
gravitational sphalerons, let us call it $S_l(\alpha)$, exists  
up to a critical value $\alpha = \alpha_s \approx 0.3095$ and no solution
of this type exists for $\alpha > \alpha_s$. However there exists
a second branch of solutions that we call $S_u(\alpha)$
for $\alpha \in [0, \alpha_s]$. 
For fixed $\alpha$ the solution
of the branch $S_u$ has a higher mass $\mu_s \equiv \mu(\infty)$
and a deeper minimum of $N(x)$
than the corresponding one on the branch $S_l$.
This is illustrated on Fig.1; the indexes $l,u$ in $S_{l,u}$ refer
to the {\it lower, upper} branch when comparing the mass.

In the limit $\alpha \rightarrow \alpha_s$, the solutions 
$S_u$ and $S_l$ converge to a common limit.  It has
\be
      \mu(\alpha_s) \approx 0.290  \ \ \ , \ \ \ 
      N_{min}(\alpha_s) \approx 0.513 
\ee
The transition
from the branch $S_l$ to the branch $S_u$ is completely smooth.

In the limit $\alpha \rightarrow 0$, the solution on $S_u$
tends to the first solution of the BM series. 
This is recovered by rescaling the radial variable $x$ according to 
$y = x/\alpha$ and taking the limit 
$\alpha = 0$, as explained in the previous section. 
The BM solution is well known but, for completeness,
we present its  profile on Fig. 2. It has
\be
     {\rm lim}_{\alpha \rightarrow \infty} \frac{\mu(\alpha)}{\alpha} 
     \approx 0.83 \ \ \ , \ \ \
     N_{min} \approx 0.242
\ee

We have studied the gravitating sphaleron for a few values of
$\epsilon$ and determined the corresponding critical value   
$\alpha_s$. These are presented in Fig. 3. The special value
$\alpha_s(0) \approx 0.376$ agrees with the result
of \cite{volkov}. We notice the rapid decrease of $\alpha_s$
for the low values of $\epsilon$.

We next discuss the solutions for a value of $\epsilon$
which allows both sphaleron and bisphaleron solutions to exist.
This occurs \cite{bk} for $\epsilon > 72.0$ and here
we choose generically $\epsilon = 100.0$ for which
the energies of the flat bisphaleron and sphaleron are
given by
\be
      E_b \approx 4.88 \ \ \ , \ \ \ E_s \approx 4.93
\ee
These two solutions get
deformed by gravity when $\alpha > 0$ and develop two
branches of solutions which we will denote $B_l(\alpha)$  
(with mass $\mu_b$) and $S_l(\alpha)$ (with mass $\mu_s$).
They exist respectively up to 
$\alpha = \alpha_b \approx 0.2247$  and 
$\alpha = \alpha_s \approx 0.2218$. At these critical values,
we have respectively
\bea
     \mu_b(\alpha_b) & \approx 0.2080 \ \ \ , \ \ \ N_{min} &\approx 0.464 \\
     \mu_s(\alpha_s) & \approx 0.2047 \ \ \ , \ \ \ N_{min} &\approx 0.463
\eea
Again, these solutions are continued by upper branches which we
denote respectively by
$B_u(\alpha)$ and $S_u(\alpha)$ and which coincide with
$B_l(\alpha)$ and with $S_l(\alpha)$ respectively at 
$\alpha = \alpha_b$ and $\alpha = \alpha_s$.
The two bisphaleron-solutions corresponding to $B_l$ and $B_u$
for $\alpha = 0.2$ are presented on Fig. 4.

All along the two branches the energy of the gravitating bisphaleron
stays slightly lower than the one of the sphaleron. Of course
both quantities become equal when $\alpha = \alpha_c$,
this is illustrated by Fig. 5.
This figure also clearly indicates that,
at the critical value $\alpha_s$ (resp. $\alpha_b$),  
the energies of
the two gravitating sphaleron $S_l, S_u$ (resp. bisphaleron $B_l, B_u$)
solutions form a cusp.
When bisphalerons are present, four solutions are
available on some interval of $\alpha$ and there are two cusps.

Completely similarly to the case $\epsilon = 0.5$, the branch $S_u$
converges to the Bartnik-Mckinnon solution in the limit $\alpha = 0$.
The scenario with the branch $B_u(\alpha)$ is different: 
decreasing $\alpha$ from $\alpha_b$ we observe that
the different radial functions composing this solution uniformly
approach their counterparts of sphaleron solution $S_u$.
This occurs for $\alpha \approx \alpha_c \approx 0.185$
as illustrated on Fig. 6. 
On this figure the quantities $f_B(\infty)$ and $H(0)$
which characterize the bisphaleron are plotted for the different branches.
The minimal value $N_{min}$ of the 
function $N(x)$ is superposed on the figure.

In view of these results we can say that the branch 
$B_u$ of solutions bifurcates from the branch $S_u$ 
at $\alpha \approx 0.185$.
 
In order to have a qualitative idea of how 
the gravitating (bi)-sphaleron
solutions behave for higher values of $\epsilon$, 
we  solved the equations for $\epsilon = 800.0$.
There the gravitating sphaleron and bisphaleron
exist respectively up to $\alpha_s \approx 0.209$
and $\alpha_b \approx 0.221$. 
The branch $B_u$ bifurcates from $S_u$ at $\alpha \approx 0.07$

This suggests that, when $\epsilon$ increases,
the branch $B_u$ bifurcates from $S_u$ for lower values 
of $\alpha$ and that
the gravitating bisphaleron exists on an interval 
which becomes relatively larger than the interval of existence of the
sphaleron.

In the limit of the non-linear sigma model
 $\epsilon = \infty$, it is known \cite{km,bk,yaffe} that the 
sphaleron is discontinuous at the origin unlike the bisphaleron
which continues to exist as a regular solution. We expect that
in this case the bisphaleron branch $B_u$ will exist up to 
$\alpha = 0$ where it will approach the Bartnik-McKinnon solution.

Let us finally say some words about the stability of the various solutions.
For $\epsilon < 72.0$ the flat sphaleron possesses a single direction of 
instability (a negative mode) and for $\epsilon > 72.0$ the sphaleron 
has two directions of instability while the bisphaleron has one \cite{yaffe,bk2}.
In this respect the sphaleron (resp. bisphaleron)
is interpreted  as  the minimal energy barrier 
\cite{km} bewteen topologically  different vacua of the Weinberg-Salam
model  for $\epsilon < 72.0$  (resp. $\epsilon > 72.0$).

The shape of the plot of the energy, with the two branches
terminating into a cusp is typical for
catastrophe theory (see e.g. \cite{kus})
and the use of  arguments based on Morse theory suggests many useful information
about the stability of the different solutions. 
For instance the number of negative modes for  the solutions on the upper branch
exceed by one unit the number of negative modes for the solutions on the lower
branch.   Such a reasonning was demonstrated
to be correct e.g. in \cite{bks}.

Using the same arguments in the present context indicates that 
the solutions on the branch $S_l$ of Fig. 1 have one direction of instability
while the solutions of the branch $S_u$ (and therefore also the BM solution) have two. 

For the solutions of Fig. 5
we have one (resp. two) directions of instability for the solutions on 
$B_l$ (resp. $B_u$)  and two  directions of instability for
the solutions on $S_l$. On the branch $S_u$ 
the number of negative modes is equal to two on the interval
$\alpha \in [0, \alpha_c]$ (i.e. before the branch $B_u$ has
 bifurcated) and equal to three for
$\alpha \in [\alpha_c, \alpha_s]$ (i.e. after the bifurcation).
The number of instable modes at the approach of the 
BM solution is then equal to two, irrespectively of the parameter
$\epsilon$.

Obviously these deductions would need to be confirmed by more elaborated calculations. 

\section{Conclusion}

The coupling of the Yang-Mills-Higgs equations to gravitation
often leads to interesting new properties of the available
gravitating classical solutions \cite{bfm1,lw}. 

The classical equations of the Weinberg-Salam model possess
a rich pattern of bifurcations of bispahleron solutions
\cite{bk,yaffe} from the sphaleron solution \cite{km}.
It is therefore natural to attempt to study the critical phenomenon
which occur in the Einstein-Weinberg-Salam equations.
In response to gravity, parametrized by the quantity $\alpha$
defined in (\ref{alpha}),
the equations exhibit another type of critical phenomenon~:
the occurence of two branches of solutions which terminate
at a critical value of $\alpha$;
this was noted in \cite{volkov} for $M_H=0$ but the phenomenon
occurs for generic values of $M_H$. 

Here we were interested only in global solutions (i.e. the
metric function $N$ has no zero on $[0,\infty]$), but
another interesting feature of gravitationally deformed
classical solutions (soliton or sphaleron) is the existence
of black hole solutions \cite{bfm1} where the function $N(r)$ develops 
an horizon at some finite value $r=r_h$, i.e. $N(r_h)=0$.
We guess that flat (bi-)sphalerons could also be deformed
in this way and produce sphaleron black holes.


\vfill
\eject

{\bf Figure captions}

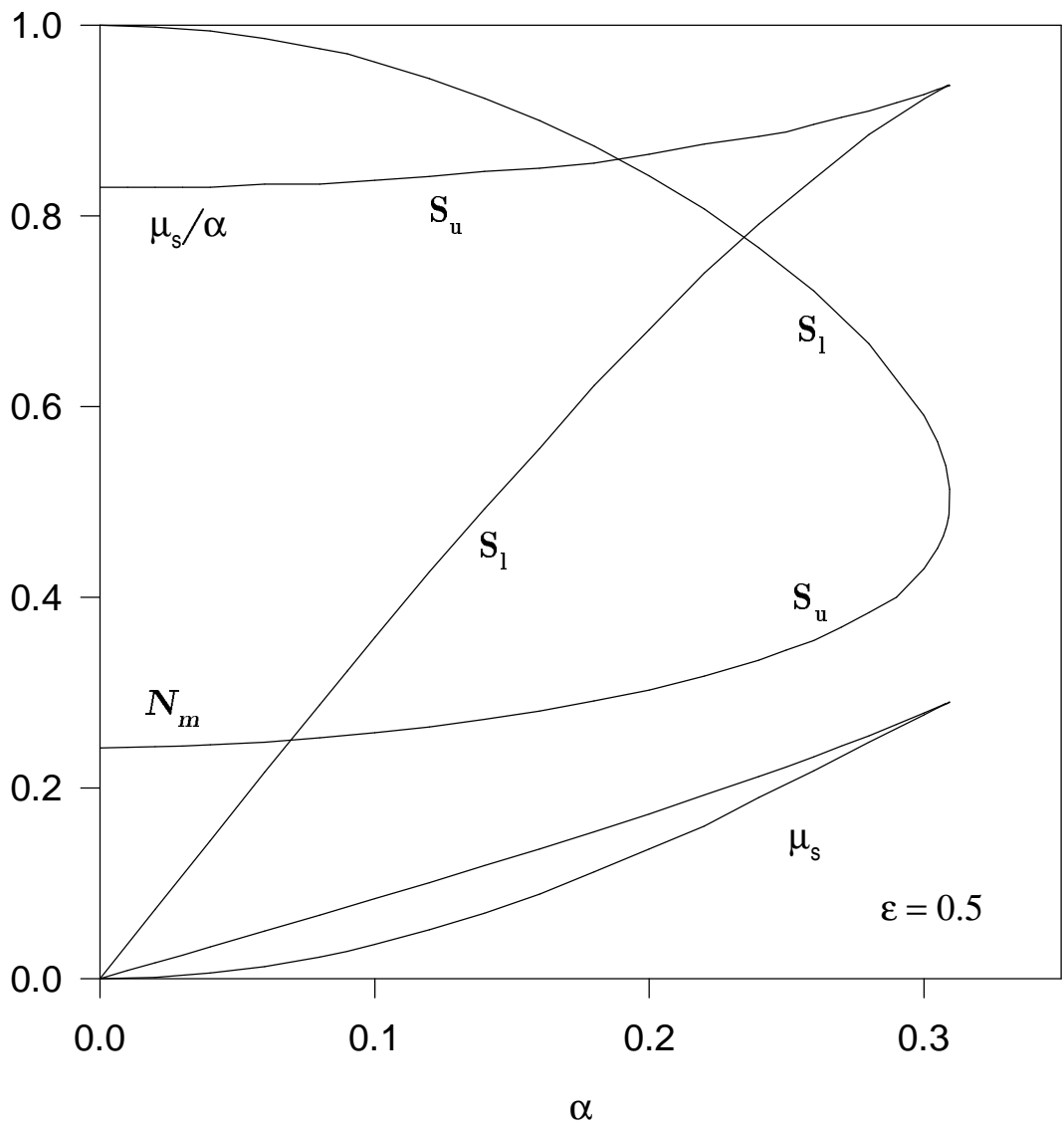
\noindent Fig.~1.

The mass of the gravitating sphaleron solutions (represented by $\mu_s$
and by  $\mu_s / \alpha$) and the minimal value $N_{m}$ of $N$ are
reported in function of $\alpha$ for $\epsilon = 0.5$.

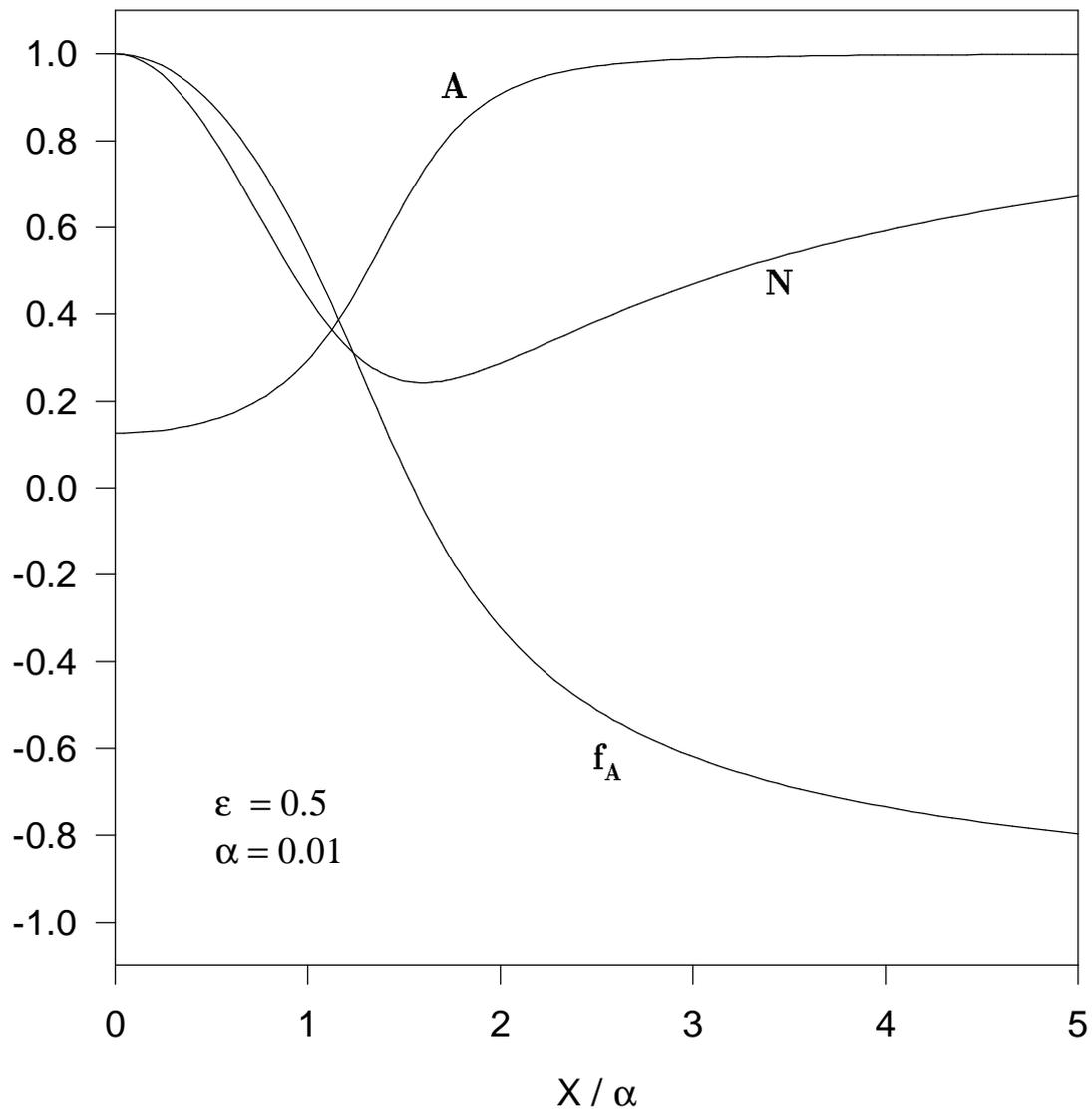
\noindent Fig.~2.

The profiles of the functions $N, A, f_A$ of the solution $S_u$
as functions of $y=x/\alpha$ for $\alpha = 0.01$ and $\epsilon = 0.5$. 

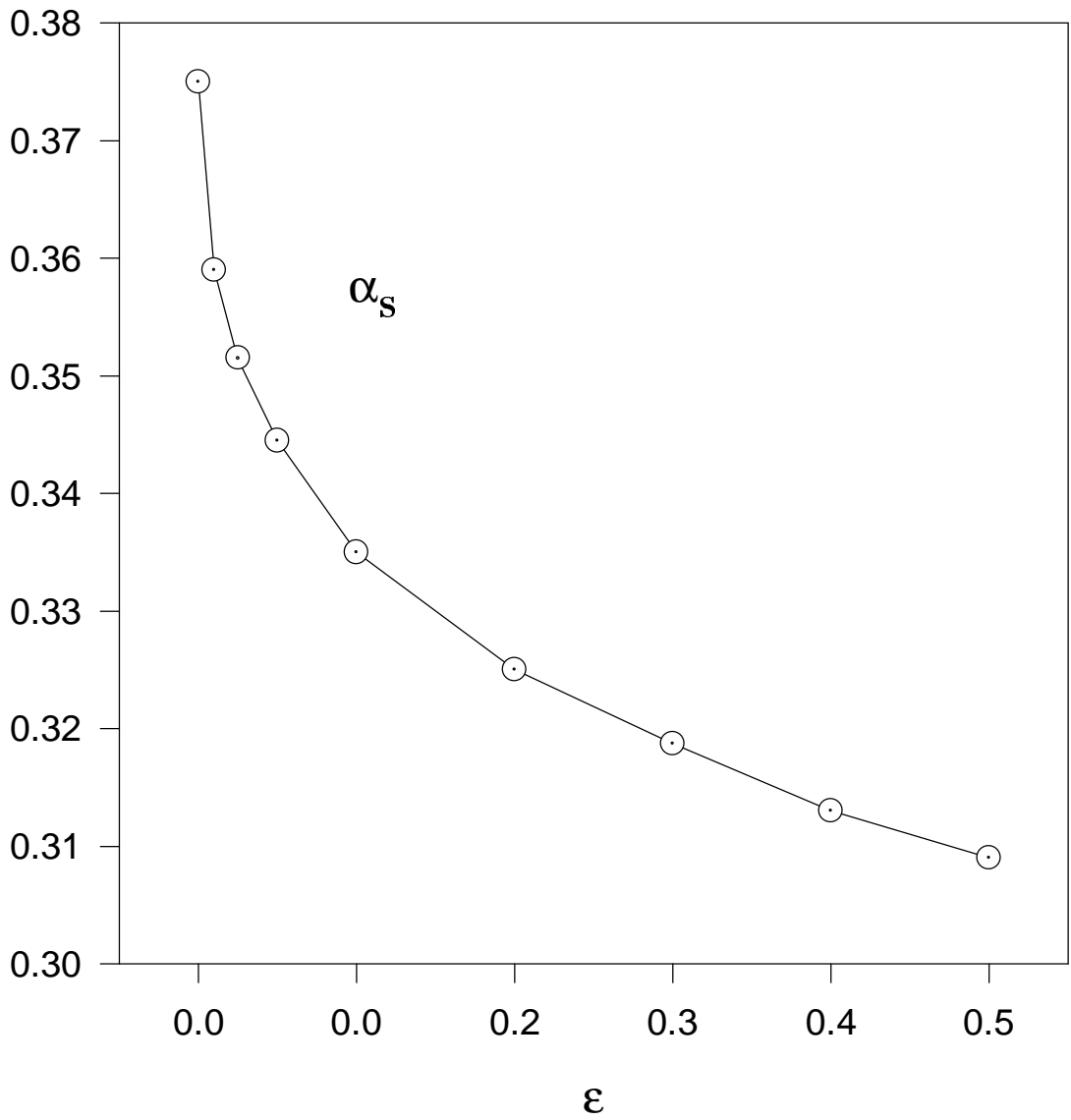
\noindent Fig.~3.

The critical value $\alpha_s$ as a function of $\epsilon$
for the low values of $\epsilon$.
 
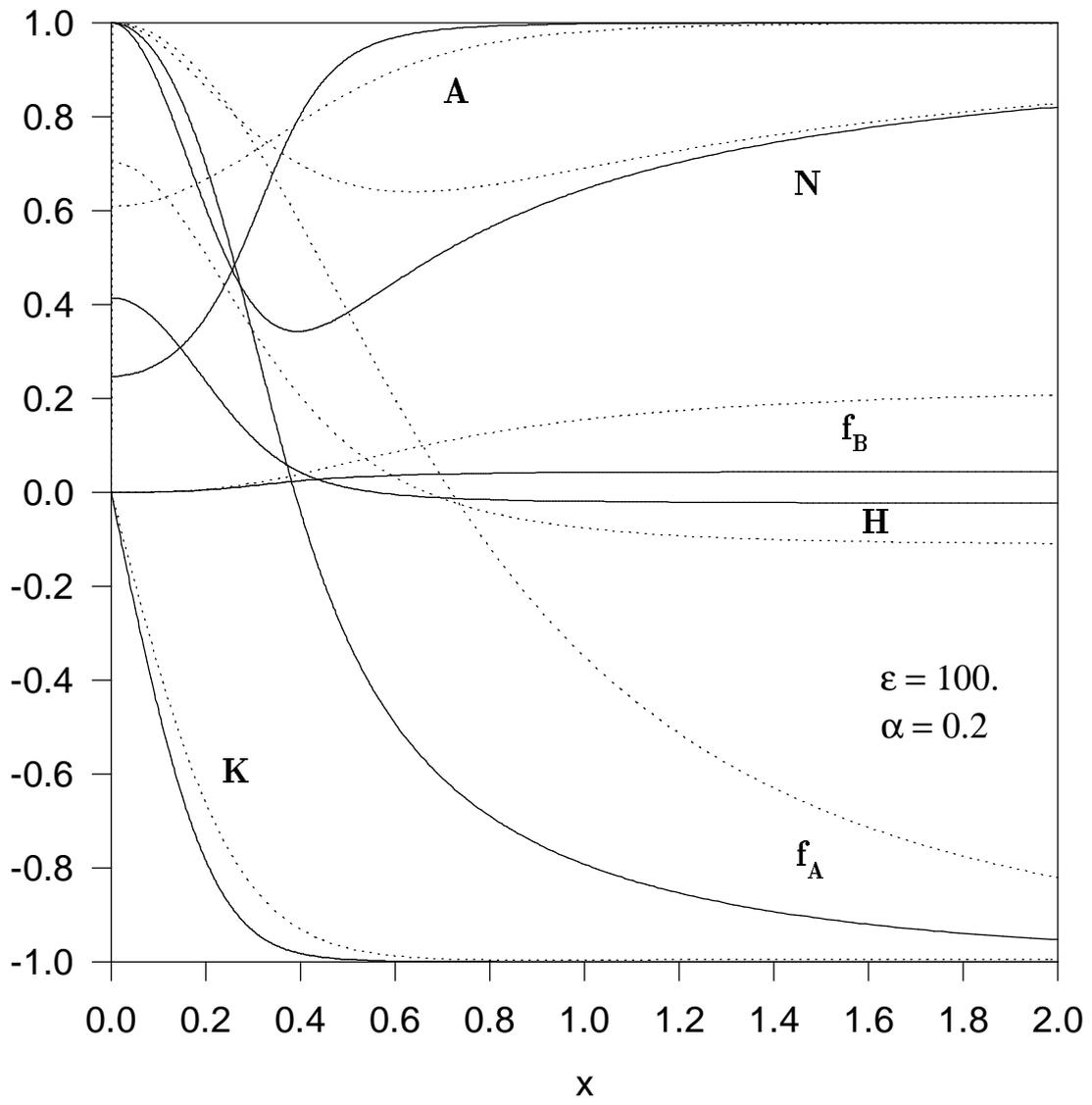
\noindent Fig.~4.

The profiles of the functions $N, A, f_A, f_B, H, K$
of the bisphalerons $B_l$ (in dotted) 
and $B_u$ (in solid) for $\alpha = 0.2$ and
$\epsilon = 100.0$. 

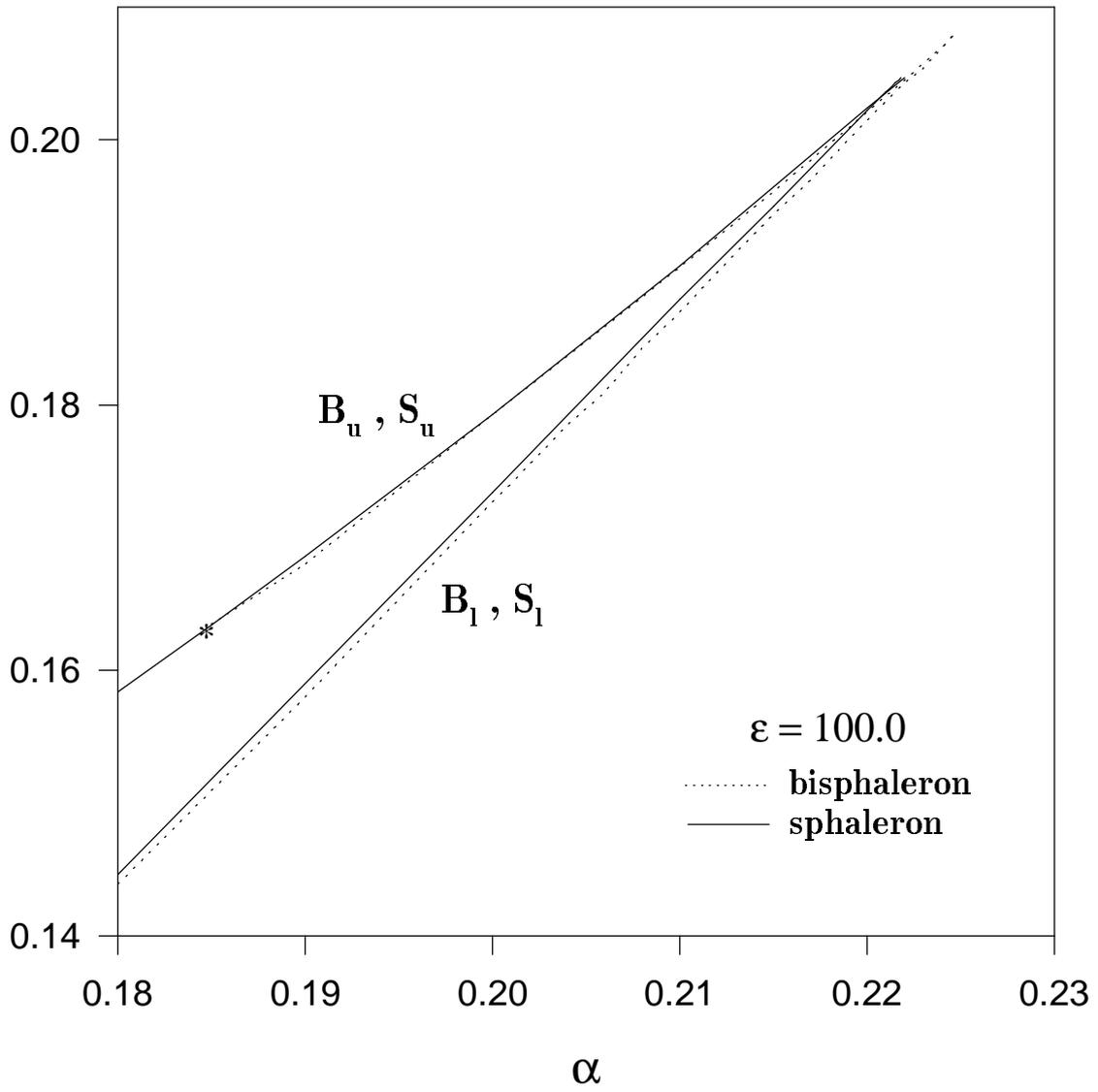
\noindent Fig.~5.

The masses $\mu_s, \mu_b$ of the gravitating sphaleron (solid) 
and bisphaleron (dotted) as functions of $\alpha$
in the region of the critical value. The branch $B_u$
stops at $\alpha \approx 0.185$, as indicated by the star.

\noindent Fig.~6.

The values of $H(0), f_{B}(\infty), N_m$ for the gravitating 
bisphaleron in function of $\alpha$.
The corresponding value of $N_{m}$ for the sphaleron
is represented by the dotted line.


\end{document}